\documentclass[a4paper]{jpconf}
\usepackage{graphicx}
\begin{document}
\title{Low-energy neutrinos}

\author{Livia Ludhova}

\address{INFN Milano, via Celoria 16, 20133 Milano, Italy \footnote{Present address: IKP-2 Forschungzentrum J\"ulich, JARA-Fame,  and RWTH Aachen, Germany}}

\ead{l.ludhova@fz-juelich.de}

\begin{abstract}
There exist several kinds of sources emitting neutrinos in the MeV energy range.  These low-energy neutrinos from different sources can be often detected by the same multipurpose detectors. The status-of-art of the field of solar neutrinos, geoneutrinos, and the search for sterile neutrino with artificial neutrino sources is provided here; other neutrino sources, as for example reactor or high-energy neutrinos, are described elsewhere. For each of these three fields, the present-day motivation and open questions, as well as the latest experimental results and future perspectives are discussed.
\end{abstract}

\section{Introduction}
\label{sec:intro}

Neutrinos interact with matter only through the weak interactions and thus, their probability to interact is small. As a consequence, neutrino detectors have to be placed in underground laboratories in which overburden rocks shield against cosmic radiations. This is the case of three experiments results of which are discussed here: Borexino at the Laboratori Nazionali del Gran Sasso in Italy and SuperKamiokande and KamLAND in the Kamioka mine in Japan.

In liquid scintillator detectors, as Borexino (280\,ton) and KamLAND (1\,kton), neutrinos are detected via the elastic scattering off electrons. Since it is not possible to distinguish the signals of such electrons from the signals due to natural radioactivity, the whole experimental setup has to be constructed from materials of very low radioactivity. Borexino, having the detection of solar neutrinos as its primary goal, has reached unprecedentedly high levels of radio-purity. Liquid scintillators provide high light yield and thus, a possibility to lower the detection threshold to sub-MeV energy region. Borexino was the first experiment to perform spectroscopic measurements of all neutrino species ({\it pp}, $^7$Be, {\it pep}, $^8$B) emitted in the {\it pp}-cycle, the chain of interactions powering the Sun by fusion of protons into $^4$He. On the other hand, the water Cherenkov detectors, as SuperKamiokande, are able to detected only higher energy neutrinos of few MeV. The possibility of directionality measurement gives an opportunity to improve the signal-to-background ratio in these detectors, by selecting only the events coming from the direction of the actual Sun's position. Thanks to its huge volume of 50 kton, SuperKamiokande is successfully collecting a large sample of $^8$B solar neutrinos.

Electron antineutrinos, as it is the case of geoneutrinos, are detected by the inverse-beta decay (IBD) reaction
\begin{equation}
\bar{\nu}_e + p \rightarrow e^+ + n.
\label{Eq:InvBeta}
\end{equation}
Liquid scintillators are used as proton-rich targets. Only antineutrinos with energies above 1.8\,MeV, the kinematic threshold of this interaction, can be detected. Reactor neutrinos, also electron flavor antineutrinos, are detected by the very same process. For an experiment constructed close to a nuclear reactor, as for example KamLAND, these represent an irreducible background in geoneutrino measurements. The IBD interaction provides, however, a powerful tool to suppress other types of background, thanks to a possibility to require a space and time coincidence between the prompt signal and the delayed one. The positron comes to rest very fast and then annihilates emitting two 511\,keV $\gamma$-rays, yielding a {\it prompt event}. The visible energy $E_{prompt}$ is directly correlated with the incident antineutrino energy $E_{\bar{\nu}_e}$: 
\begin{equation}
E_{prompt}= E_{\bar{\nu}_e}- 0.784 \rm ~{MeV}.  
\label{Epro}
\end{equation}
The neutron, produced in IBD reaction together with prositron, keeps initially the information about the incident ${\bar{\nu}_e}$ direction. Unfortunately, it is typically captured on protons only after a long thermalization time with $\tau$ = 200 - 250\,$\mu$s (depending on scintillator), during which this information is mostly lost. When the thermalized neutron is captured on proton, a 2.22\,MeV de-excitation $\gamma$-ray is emitted, providing a coincident {\it delayed event}.  

After this short introduction to the experimental techniques of low-energy neutrino physics, the following three sections are dedicated to solar neutrinos, geoneutrinos, and neutrinos from artificial radioactive sources, respectively.  Each of them describes the present-day motivation of each particular field, its open questions, as well as the latest experimental results and future perspectives.

\section{Solar neutrinos}
\label{sec:solar}

 \begin{figure}[t]
\includegraphics[width=20pc]{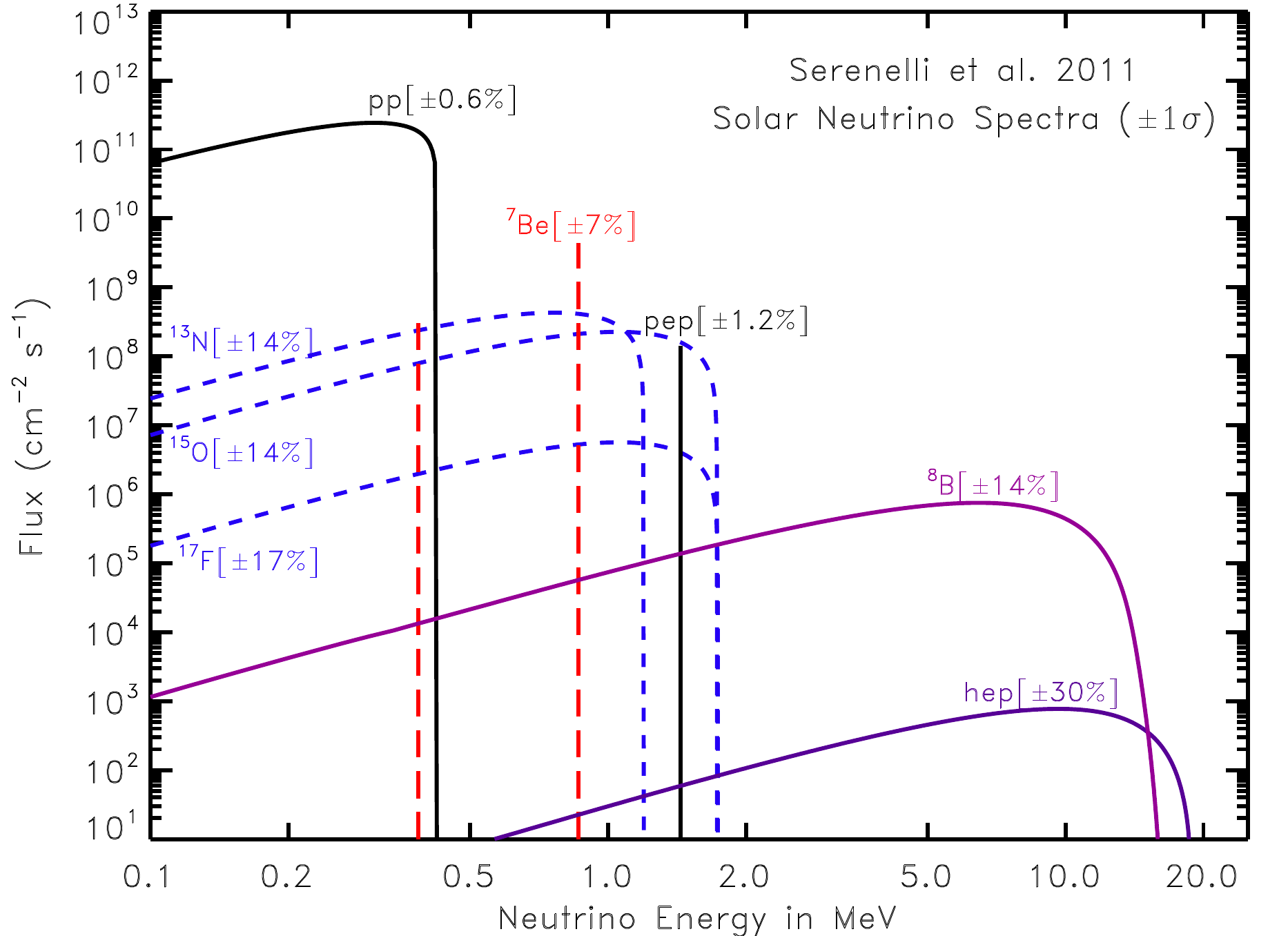}\hspace{2pc}%
\begin{minipage}[b]{0.40\textwidth}
\caption{The solar neutrino spectrum predicted by the Standard Solar Model calculation of~\cite{Serenelli2011}.}
\label{Fig:solar_spectrum}
\end{minipage}
\end{figure}
 
The Sun is powered by nuclear fusion reactions in which protons are forming $^4$He nuclei. The principal chain of reactions of this process is called the {\it pp}-cycle. Figure~\ref{Fig:solar_spectrum} shows the energy spectrum of neutrinos emitted in some of these reactions. The so called {\it pp} neutrinos represent the dominant component with continuous energy spectrum up to 420\,keV. There are 3 mono-energetic lines, the main $^7$Be line at 862\,keV and the one at 384\,keV, with the intensity $\sim$10-times lower, and the low-intensity {\it pep} neutrino line at 1440\,keV. Only the liquid scintillator experiments are able to perform the spectroscopy of these species of solar neutrinos. The water Cherenkov based detectors are instead able to measure only the higher energy part of $^8$B neutrinos, extending up to $\sim$15\,MeV and having a very low flux. It is believed that a small fraction of solar energy is produced in the fusion process catalyzed by the presence of heavier elements, in the so called CNO cycle, which should be the dominant fusion process in heavy stars. The CNO-neutrinos, shown in Fig.~\ref{Fig:solar_spectrum} by the dashed blue lines labeled $^{13}$N, $^{15}$O, and $^{17}$F, have never been observed yet.

The measurement of solar neutrinos brought a real breakthrough in particle physics through the observation of neutrino oscillations~\cite{SNOOscill}, proving that neutrinos have non-zero mass. This discovery made in 2002 was awarded the Nobel prize now in 2015. And even today, the field of solar neutrinos continues to be a thrilling research area giving us a possibility of new discoveries.

 \begin{figure}[t]
\includegraphics[width=20pc]{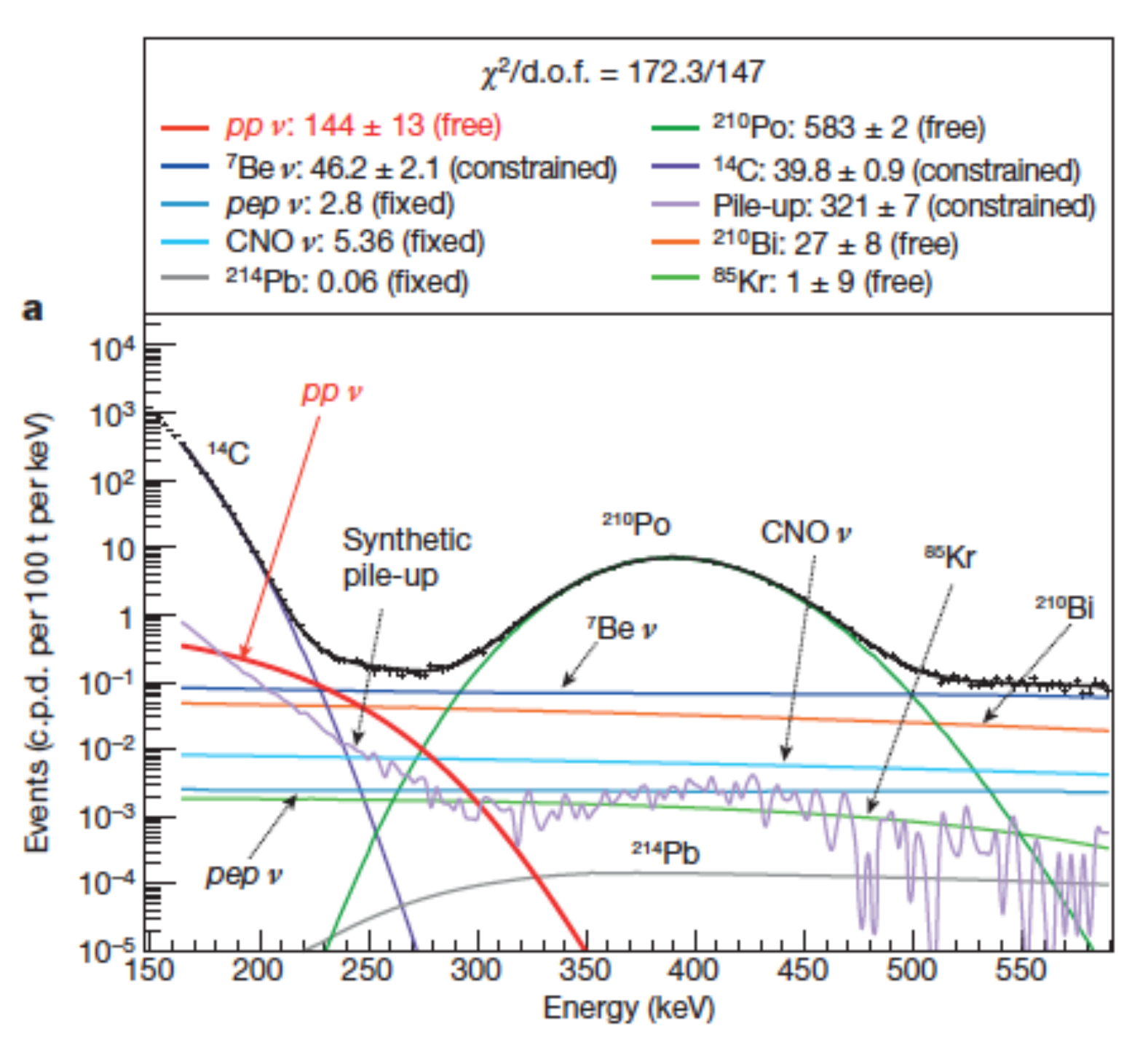}\hspace{1pc}%
\begin{minipage}[b]{0.40\textwidth}
\caption{\label{label} Borexino data with spectral fit leading to the extraction of {\it pp}-neutrino interaction rate~\cite{Borexinopp}. The observed rate of {\it pp} neutrinos is $(144 \pm 12 {\rm (stat)} \pm 10 {\rm (sys)}$\,cpd/100 ton, corresponding to flux of $6.6 \times (1 \pm 0.106) \times 10^{10}$ cm$^{-2}$ s$^{-1}$.}
\label{Fig:Borexinopp}
\end{minipage}
\end{figure}

 \begin{figure}[t]
\includegraphics[width=21pc]{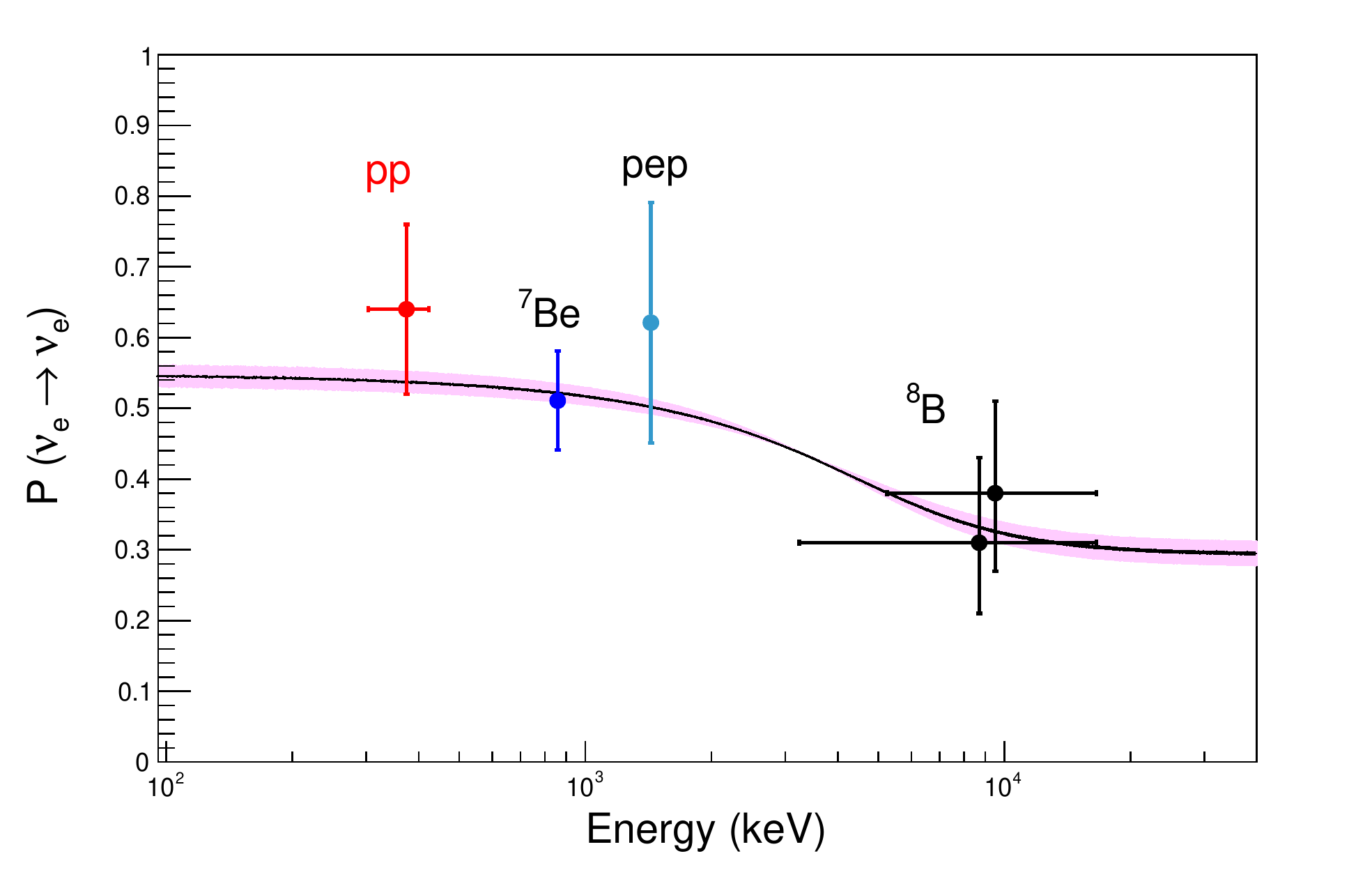}\hspace{1pc}%
\begin{minipage}[b]{0.40\textwidth}
\caption{Electron-neutrino survival probability for solar neutrino species as measured by Borexino~\cite{Borexinopp} compared to the prediction of LMA-MSW solution of neutrino oscillations~\cite{MSW}.}
\label{Fig:BorexinoPee}
\end{minipage}
\end{figure}

Solar neutrinos can help us to understand the Sun itself. The so called Standard Solar Models (SSM) have among other inputs (luminosity, mass, radius, opacity etc.) also the so called {\it metallicity}, the ratio of abundances of elements heavier than He (C, N, O, Ne, Mg, Si, Ar, Fe) to that of hydrogen. Among SSM outputs are the neutrino fluxes (as in Fig.~\ref{Fig:solar_spectrum}) and the sound-waves speed profiles (helioseismology). The hot topic in solar neutrino physics today is the so called metallicity problem: the new 3D SSM~\cite{highZ} using new spectroscopic measurements with {\it high metallicity} do spoil the perfect agreement between the helioseismological data and the predictions of the old 1D SSM~\cite{lowZ} using in input the older spectroscopic data with {\it low metallicity}. Since the low- and high-metallicity SSM predict different fluxes of neutrinos (difference of 8.8, 17.7, and 30-40\% for $^7$Be, $^8$B, and CNO neutrino fluxes, respectively), the precise measurements of solar neutrino fluxes can significantly help in solving this puzzle. Unfortunately, the existing measurements of $^7$Be and  $^8$B fluxes fall in between the predictions of the high- and low-metallicity SSM, being compatible with both of them (see Fig. 82 in~\cite{BXLong}).

Neutrinos are in fact the only direct evidence of the ongoing fusion interactions in the Sun's core. By comparing the neutrino luminosity with the optical one, one can test how well do we understand these mechanisms powering our star.  It takes at the order of $10^5$ years before the photon produced in the Sun's core can reach the Earth, while the only-weakly-interacting neutrino can immediately escape the Sun and reach us in few minutes. Thus, by comparing the optical and neutrino luminosities, one can study also the Sun's stability at the time scale of  $10^5$ years. In this sense, the spectral observation of {\it pp} neutrinos by Borexino~\cite{Borexinopp}, shown in Fig.~\ref{Fig:Borexinopp}, was ranked by Physics World  among the Top 10 Physics Breakthroughs of 2014.

 \begin{figure}[t]
\includegraphics[width=20pc]{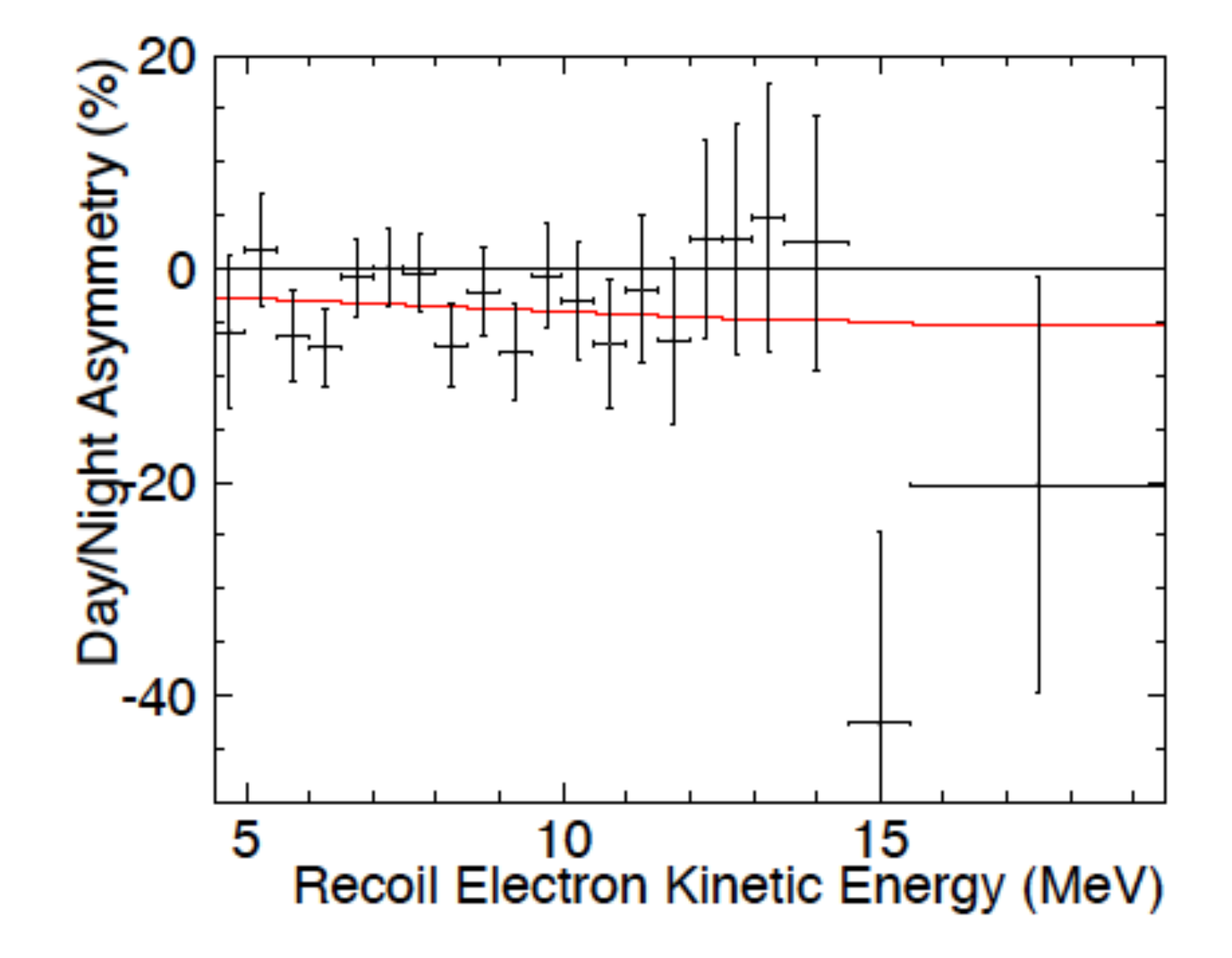}\hspace{1pc}%
\begin{minipage}[b]{0.40\textwidth}
\caption{Day/Night asymmetry of $^8$B solar neutrinos rate as a function of recoil electon energy measured by SuperKamiokande~\cite{SuperKDN}}
\label{Fig:SuperKDN}
\end{minipage}
\end{figure}

Solar neutrino studies can also contribute to our understanding of the nature of neutrino interactions. The electron-neutrino survival probability $P_{ee}$ as a function of energy, as predicted by the LMA-MSW solution of neutrino oscillations~\cite{MSW}, is demonstrated by the curve in Fig.~\ref{Fig:BorexinoPee}. The matter resonance effect, important especially in the dense solar core, decreases the electron-neutrino survival probability at energies above cca 5\,MeV (only $^8$B solar neutrinos). The data points for {\it pp}, $^7$Be, {\it pep}, and $^8$B neutrinos measured by Borexino and shown in Fig.~\ref{Fig:BorexinoPee} are pinning down the predicted theoretical curve. However, the large volume Cherenkov detectors, as SuperKamiokande, are better suited to further test the so called transition region where the survival probability curve changes from the vacuum- to matter-dominated region~\cite{SuperK_paralel,SuperKSolar}. This is particularly important in a view of the fact that some new physics could influence the exact shape of this transition region, as for example the Non-Standard neutrino Interactions (NSI)~\cite{NSI} or 
 the existence of an ultra-light sterile neutrino~\cite{LightNu}.

 \begin{figure}[b]
\includegraphics[width=20pc]{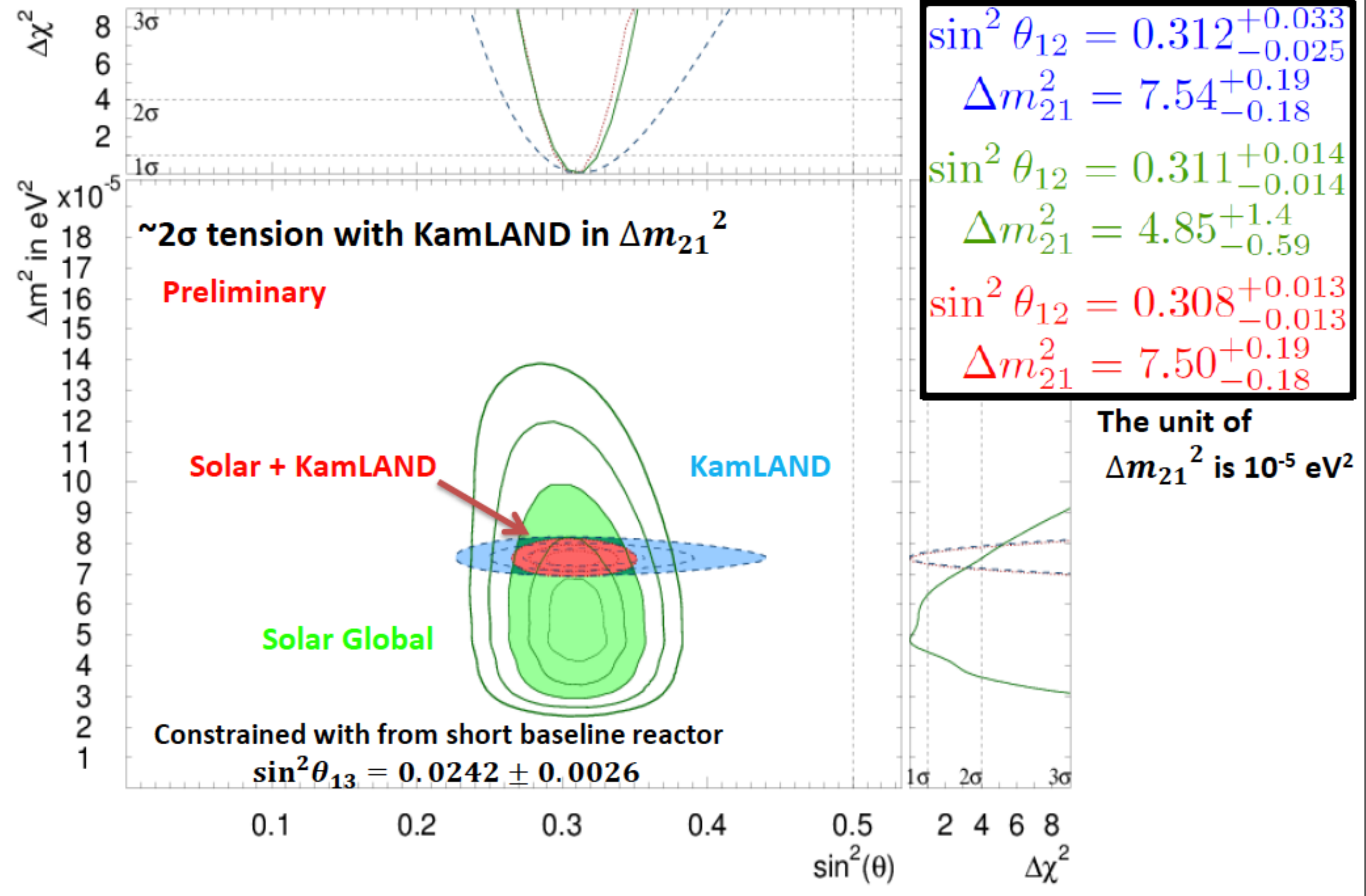}\hspace{1pc}%
\begin{minipage}[b]{0.40\textwidth}
\caption{Allowed contours of $\Delta m^2_{21}$ versus $\sin \theta_{12}$ from solar neutrino data (green) at 1, 2, 3, 4, and 5$\sigma$ and KamLAND reactor antineutrino data (blue) at 1, 2, and 3$\sigma$ confidence levels. Combined result is shown in red. Taken from~\cite{SuperK_paralel}.}
\label{Fig:OscilPar}
\end{minipage}
\end{figure}

The matter resonance effect can cause also the regeneration of the electron flavor as solar neutrinos cross the Earth during the night. MSW-LMA predicts this effect to be energy dependent. In accordance, Borexino has excluded this night regeneration for $^7$Be neutrinos with about 1\% precision~\cite{BorexinoDN}. However, for higher energies of $^8$B neutrinos, SuperKamiokande succeeded to provide the first direct indication at 3$\sigma$ confidence level of such matter enhanced oscillations~\cite{SuperKDN}, as demonstrated in Fig.~\ref{Fig:SuperKDN}.

Global fits of all solar neutrino data result in the extraction of neutrino oscillation parameters, being most sensitive to $\sin^2\theta_{12}$ and $\Delta m^2_{12}$. Figure~\ref{Fig:OscilPar} demonstrates the latest update provided by SuperKamiokande~\cite{SuperK_paralel} for  $\sin^2\theta_{12}$ and $\Delta m^2_{12}$ parameter space. A 2$\sigma$ confidence level tension is observed for the $\Delta m^2_{12}$ value obtained from solar neutrino data and that coming from KamLAND measurements with reactor antineutrinos~\cite{Kamland}.
 
Borexino is currently running in the so called Phase II, after an extensive purification campaign.  More precise measurements of the $^7$Be, {\it pep} and may be also {\it pp} neutrino fluxes can be expected in near future. A strong effort is invested also towards a possible observation of CNO neutrinos, an extremely difficult task due to the spectral degeneracy with $^{210}$Bi. SuperKamiokande continues to take data and to further test the $P_{ee}$ transition region. Other future large volume liquid scintillator detectors (1\,kton SNO+ in Canada should start soon and 20\,kton JUNO in China has its start planned for 2020) have solar neutrinos among their additional scientific goals.

 \section{Geoneutrinos}
\label{sec:geo}

 \begin{figure}[t]
\begin{centering}
\includegraphics[width=27pc]{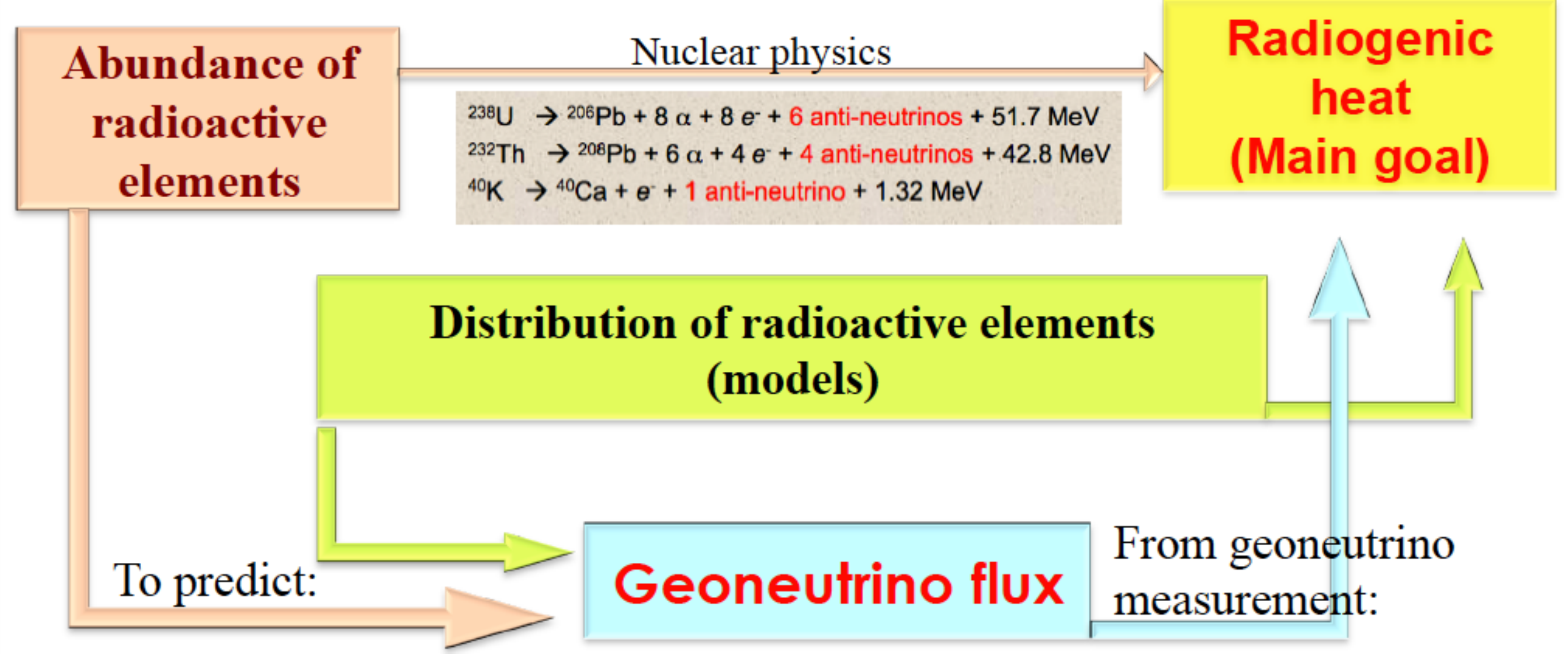}
\caption{Schematic visualization of the relations among the abundances and distributions of radioactive elements inside the Earth with the geoneutrino flux and radiogenic heat.}
\label{Fig:Geo}
\end{centering}
\end{figure}

Geoneutrinos are electron-flavor antineutrinos emitted in the $\beta$ decays of long-lived radioactive elements, called also {\it the heat producing elements} (HPE). As shown in schematic reactions in Fig.~\ref{Fig:Geo}, geoneutrinos are emitted along the decay chains of $^{238}$U and $^{232}$Th and in the $^{40}$K decay. The main aim of geoneutrino studies is to determine the Earth's radiogenic heat, especially the unknown contribution from the mantle. The mantle composition is quite unknown with respect to the better-known crustal composition. Knowing the mass/abundances of HPE, the radiogenic heat is directly determined. The geoneutrino studies are, however, complicated through an unknown distribution of HPE, on which depends both the  geoneutrino signal prediction as well as the final interpretation of the measured geoneutrino flux, see Fig.~\ref{Fig:Geo}.

 \begin{figure}[t]
\includegraphics[width=20pc]{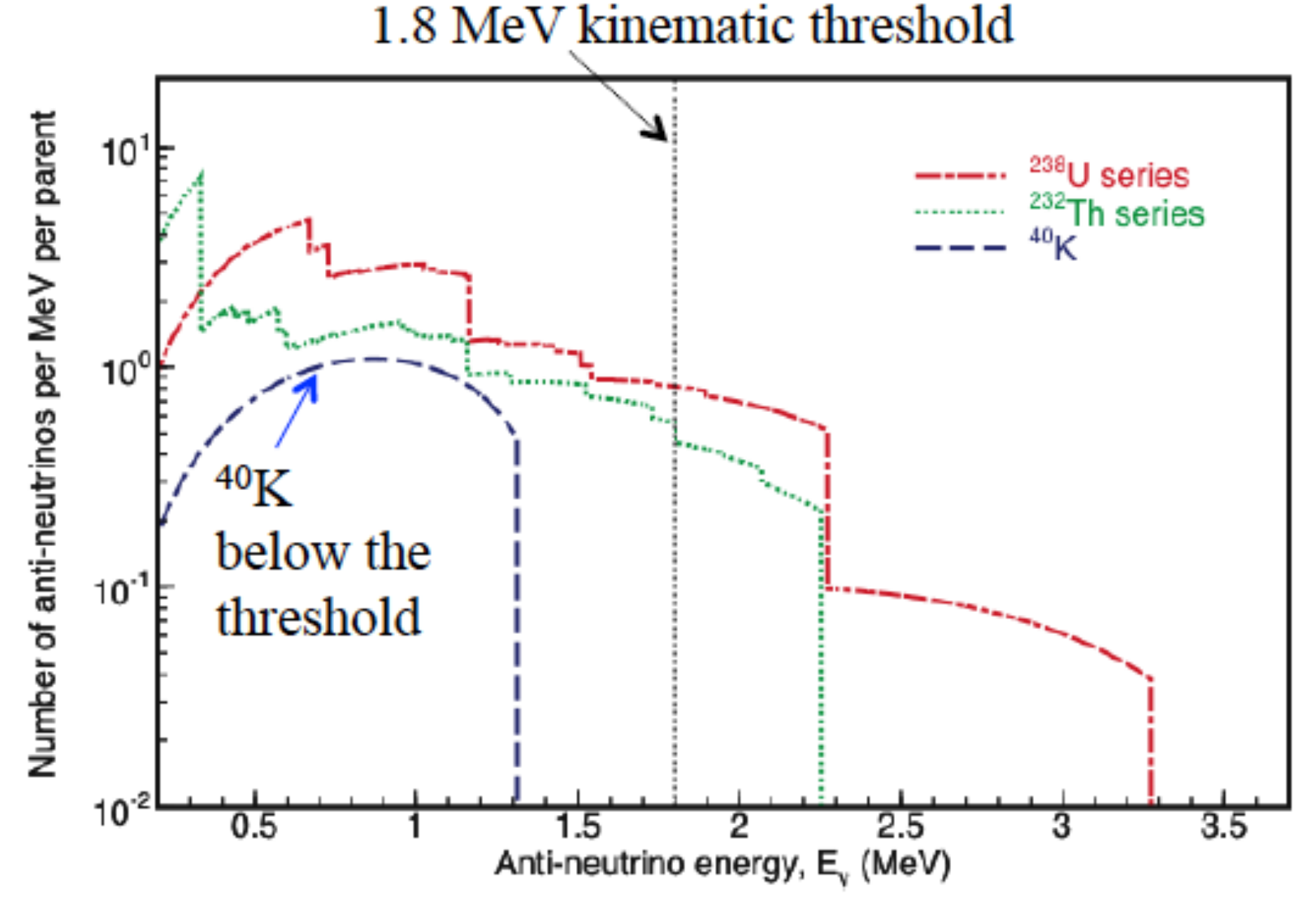}\hspace{1pc}%
\begin{minipage}[b]{0.40\textwidth}
\caption{Energy spectrum of geoneutrinos. The dashed vertical line shows the energy threshold of the IBD interacition.}
\label{Fig:GeoSpectrum}
\end{minipage}
\end{figure}

Geoneutrino spectrum is shown in Fig.~\ref{Fig:GeoSpectrum}.  Due to the 1.8\,MeV kinematic threshold of the IBD reaction (Eq.~\ref{Eq:InvBeta}), only the high-energy tail of $^{238}$U and $^{232}$Th geoneutrinos can be detected, while all $^{40}$K geoneutrinos are below the threshold. 

 \begin{figure}[t]
\includegraphics[width=20pc]{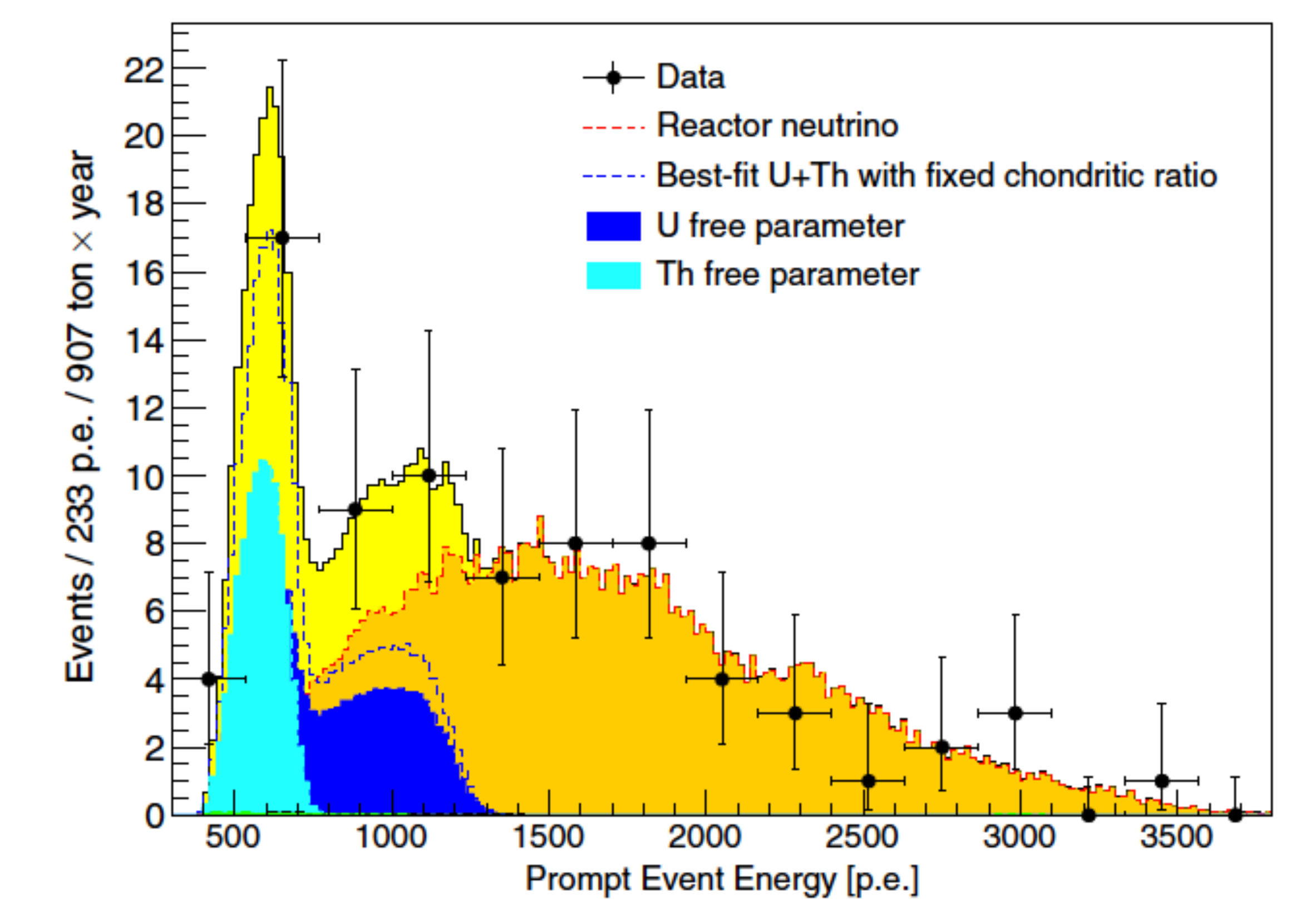}\hspace{1pc}%
\begin{minipage}[b]{0.40\textwidth}
\caption{Prompt light yield spectrum in photoelectrons (p.e.) of 77 antineutrino candidates measured by Borexino and the best fit~\cite{GeoBx2015}. 1\,MeV corresponds to $\sim$500\,p.e.}
\label{Fig:GeoBx}
\end{minipage}
\end{figure}

Uranium and Thorium are so called refractory litophile elements. This means, that {\it i)} they do bound to silicates and not to metals (so they are not expected to be present in the Earth's metallic core) and {\it ii)} in the process of partial melting of the heterogeneous system as are the rocks, they do concentrate in partial melt which is typically lighter than the HPE-depleted solid residuum. The latter characteristics means that the HPE are strongly concentrated in the continental crust which has the most complex geological history. The oceanic crust is enriched in HPE with respect to the mantle, from which it is differentiated along mid-ocean ridges. Typically, for the experimental sites built at the continental crust, about half of the total geoneutrino signal comes from the continental crust in the area of few hundreds of kilometers around the detector~\cite{Huang}. Thus, for these experiments, in order to extract the mantle contribution from the measured geoneutrinos signal, it is necessary to be able to subtract the crustal contribution. This means, that the local geology in the area of the experiment has to be known.

Geoneutrinos represent a completely new technique how to get information about the deep Earth. Seismology, by studying the propagation of the $S$ and $P$ waves, can reconstruct the velocity and the density profiles of the Earth, but does not provide any direct information about the chemical composition. This in turn is the topic of study for geochemistry. Direct rock samples are available only from limited depths and for the deep Earth we have to rely on geochemical modeling. This uses the correlations observed between the compositions of meteorites and solar photosphere~\cite{GeoRev}, which are then extrapolated also for the Earth. The bulk composition of the silicate Earth, the so called Bulk Silicate Earth (BSE) models (for their overview see~\cite{Sramek}), describe the composition of the Primitive Mantle, after the core separation and before the crust-mantle differentiation. Geodynamical BSE models, considering the mantle convection, require in general higher abundances of HPE in the mantle with respect to geo- and cosmo-chemical BSE models. Information about the integral surface heat flux of 47 $\pm$ 2\,TW~\cite{Davis} is derived from the temperature gradients measured along the bore-holes.

Today, only two experiments succeeded to measure geoneutrinos: KamLAND in Japan and Borexino in Italy. The latest KamLAND result, $116 ^{+28}_{-27}$ geoneutrinos detected with $4.9 \times 10^{32}$ target-proton $\times$ year exposure,  is from 2013~\cite{GeoKL}, including the period of low reactor antineutrino background after the April 2011 Fukushima disaster. 

Borexino provided a new update in 2015~\cite{GeoBx2015}, as demonstrated in Fig.~\ref{Fig:GeoBx}. Within the exposure of $(5.5 \pm 0.3)  \times 10^{31}$ target-proton $\times$ year, $23.7 ^{+6.5}_{-5.7} \rm{(stat)} ^{+0.9}_{-0.6} \rm{(sys)} $ geoneutrino events have been detected. The null observation of geoneutrinos has a probability of $3.6 \times 10^{-9}$ (5.9$\sigma$). A geoneutrino signal from the mantle is obtained at 98\% confidence level. The radiogenic heat production for U and Th from the present best-fit result is restricted to the
range 23 to 36 TW, taking into account the uncertainty on the distribution of HPE inside the Earth.

In the near future, a possible update could be expected from KamLAND, possibly after the refurbishing of the outer detector planned for 2016. Borexino geoneutrino data-set will naturally end with the start of the SOX project (see below) at the end of 2016. Future experiments SNO+ and JUNO have geoneutrinos among their scientific goals. A real breakthrough would come with the proposed Hanohano~\cite{hanohano} project in Hawaii.  Placed underwater on a thin, HPE depleted oceanic crust, the mantle contribution to the total geoneutrino flux would be dominant.

\section{Neutrinos from artificial radioactive sources}
\label{sec:sources}

The measurement of decays of Z0 boson at LEP has confirmed that there are 3 families of light neutrinos interacting through weak interactions. In recent years, however, several experimental results indicate that the 3-flavor picture might not be complete: deficit observed in reactor antineutrino short-baseline experiments, the so called reactor anomaly~\cite{reactorAnomaly}, the deficit observed in calibration of Gallex and SAGE experiments with neutrino sources~\cite{CalibAnomaly}, the famous appearance of ${\bar \nu}_e$ in ${\bar \nu}_\mu$ beam seen by LSND~\cite{LSND}  and further tested by MiniBooNe~\cite{MiniBooNE}. All these anomalies could be at least partially accommodated if there would exist at least one sterile neutrino. It is called sterile because it would not interact through the forces described by the Standard Model. A global analysis of short-baseline neutrino oscillation data~\cite{Giunti} in 3+1 parameter space leads to $\Delta m^2_{41}$ between 0.82 and 2.19 eV$^2$ at 3$\sigma$. In order to probe this region with MeV (anti)neutrinos, the oscillation experiments at distances of few meters have to performed. A possible realization of such measurements is to place a very strong (anti)neutrino source just next to a large volume neutrino detector. In such an experiment, the hypothetical existence of sterile neutrino could be observed not only through an absolute disappearance but also through the oscillation pattern in (L, E) parameter space deviating from the standard 3-flavor scenario.
Such an observation would be a “smoking gun” prove that the cause of the deficit is the phenomenon of neutrino oscillation into a new sterile flavor.

There are several candidate projects around the world. However, the only project which has a planned start of the data acquisition in 2016 is Short-distance neutrino Oscillation with BoreXino (SOX) in Italy~\cite{SOX}. A 3.7 PBq (100 kCi) $^{144}$Ce-$^{144}$Pr antineutrino source, producing about $10^{15}$ ${\bar \nu}_e /$ s, will be placed below the Borexino detector at the end of 2016. The source distance from the detector center will be 8.3\,m. The employment of such a strong source is both a technical and a bureaucratic challenge. $^{144}$Ce with 285 day lifetime decays to $^{144}$Pr which then with much shorter lifetime decays to $^{144}$Nd. SOX will measure antineutrinos above the IBD threshold, emitted by $^{144}$Pr with the end-point at about 3\,MeV. The measurement is complicated by the emission of the 2.2\,MeV gammas from the de-excitation of $^{144}$Nd. The source will be produced by Mayak in Russia by extracting $^{144}$Ce from the fresh nuclear spent fuel.

 \begin{figure}[t]
\includegraphics[width=18pc]{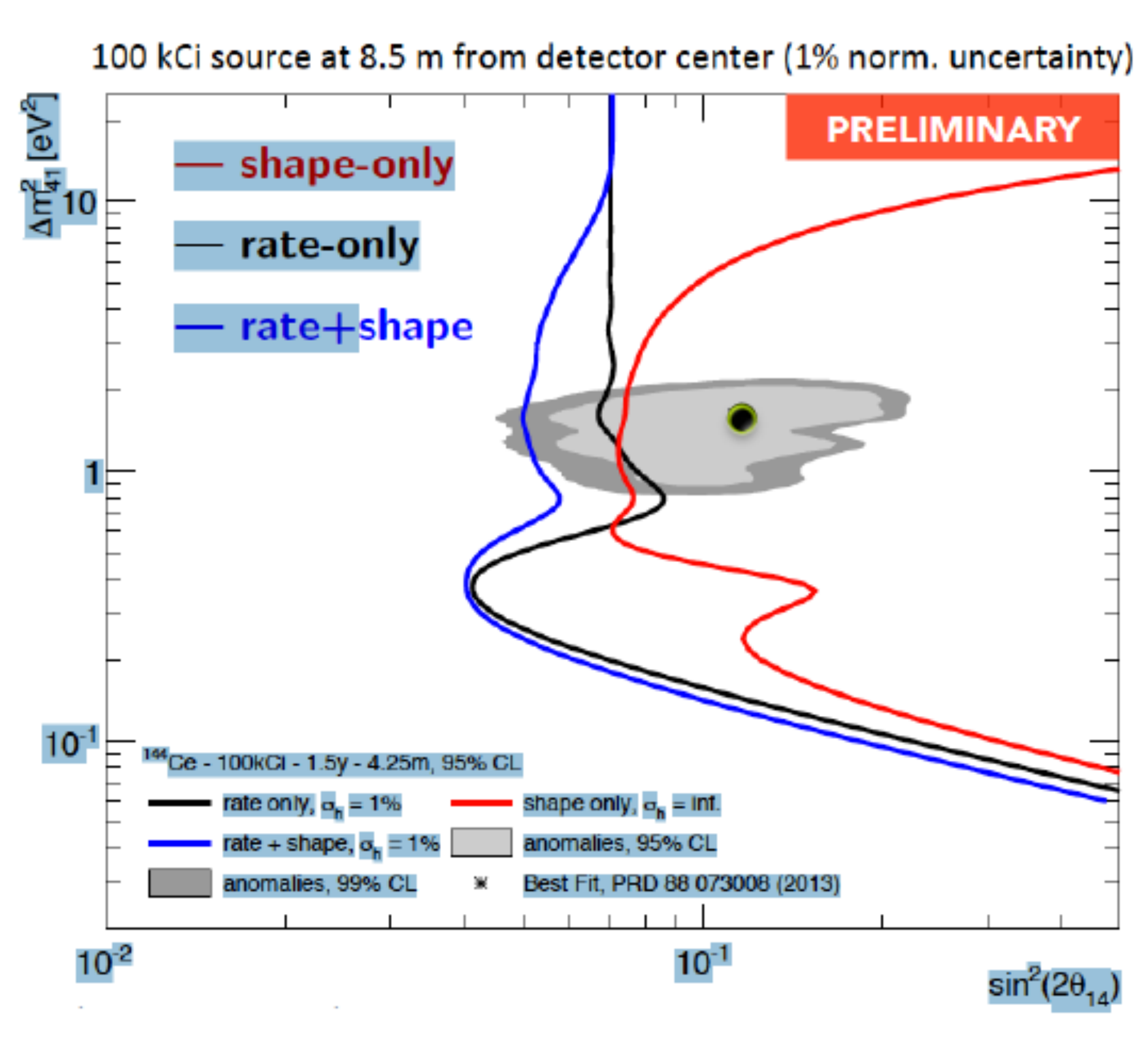}\hspace{1pc}%
\begin{minipage}[b]{0.40\textwidth}
\caption{Expected sensitivity of the SOX project to sterile neutrino parameters $\Delta m^2 _{41}$ and $\sin^2(2\theta_{14})$~\cite{SOX_paralel}.}
\label{Fig:SOX}
\end{minipage}
\end{figure}

This measurement is facing several additional complications. The $^{144}$Ce-$^{144}$Pr spectra have to be known with 5\% precision, but the situation is complicated since we are dealing with   forbidden transitions. The development of new precise spectroscopic measurements is ongoing. The absolute activity of the source will be measured through calorimetric measurement with about 1.5\% precision. An extensive calibration campaign of the Borexino detector with AmBe neutron source and other gamma radioactive sources is planned for 2015. The aim is the determination of the detector response function in the whole scintillator volume with 1 to 2\% precision. The expected preliminary sensitivity to sterile neutrino parameters $\Delta m^2 _{41}$ and $\sin^2(2\theta_{14})$ is shown in Fig.~\ref{Fig:SOX}.


\section*{References}
\begin{thebibliography}{16}
\bibitem{Serenelli2011} Serenelli A, Haxton W C and  Pena-Garay C 2011 {\it Ap. J.} {\bf 743}  24 
\bibitem{SNOOscill} Ahmad Q R et al. (SNO Collaboration) 2002 {\it Phys. Rev. Lett.} {\bf 89} 011301-1
\bibitem{Borexinopp} Bellini G et al. (Borexino Collaboration) 2014 {\it Nature} {\bf 512}  383
\bibitem{highZ} Serenelli A M, Haxton W vC and Pe\~na-Garay C 2011 {\it Ap. J.} {\bf 743}  24
\bibitem{lowZ} Asplund M, Basu S, Ferguson J W and Asplund M 2009 {\it Ap. J. Lett.} {\bf 705} L123
\bibitem{BXLong}  Bellini G et al. (Borexino Collaboration)  2014 {\it Phys. Rev. D} {\bf 89} 112007
\bibitem{MSW} Mikheyev S P and Smirnov A Yu 1985 {\it Sov. J. Nucl. Phys.} {\bf 42} 913; Wolfenstein L (1978) {\it Phys. Rev. D} {\bf 17} 2369 
\bibitem{SuperK_paralel} Nakano Y for SuperKamiokande Collaboration: Solar neutrino results from SuperKamiokande. Parallel talk at TAUP 2015, Torino, Italy.
\bibitem{SuperKSolar} Renshaw A for SuperKamiokande Collaboration 2014 {\it Phys. Proc.} {\bf 00} 1
\bibitem{NSI} Minakata Y and Pe\~na-Garay C (2012)  {\it Adv. High En. Phys.} {\bf 2012} ID 349686
\bibitem{LightNu} de Holanda P C and Smirnov A Yu (2004) {\it Phys. Rev. D} {\bf 69} 113002; de Holanda P C and Smirnov A Yu (2011) {\it Phys. Rev.} {\bf 83} 113011
\bibitem{BorexinoDN} Bellini G et al. (Borexino Collaboration)  2012 {\it Phys. Lett. B} {\bf 707}  22
\bibitem{SuperKDN} Renshaw A et al. (SuperKamiokande Collaboration)  arXiv:1312.5176
\bibitem{Kamland} Abe S et al. (KamLAND Collaboration) 2008 {\it Phys. Rev. Lett. } {\bf 100}  221803
\bibitem{Huang} Huang Y at al. 2013 {\it Geochem. Geophys. Geosyst.} {\bf 14} 2003
\bibitem{GeoRev} Bellini G et al. 2013 {\it Prog. Part. Nucl. Phys.} {\bf 73} 1
\bibitem{Sramek}  \v{S}r\'amek et al. 2013 {\it Earth and Plan. Sci. Let.} {\bf 361} 356
\bibitem{Davis} Davies J H and Davies D R  2010 {\it Solid Earth} {\bf  1}  5
\bibitem{GeoKL} Gando A et al. (KamLAND Collaboration) 2013 {\it Phys. Rev. D} {\bf 88} 033001
\bibitem{GeoBx2015} Agostini M et al. (Borexino Collaboration) 2015 {\it Phys. Rev. D} {\bf 92} 031101 (R)
\bibitem{hanohano} Learned J G, Dye S T and Pakvasa S 2008 arXiv: 0810.4975
\bibitem{reactorAnomaly} Mention G et al. (2011) {\it Phys. Rev. D} {\bf 83} 073006; Mueller A et al. (2011) {\it Phys. Rev. C} {\bf 83} 054615
\bibitem{CalibAnomaly} Giunti C and Laveder M 2011 {\it Phys. Rev. C} {\bf 83} 065504
\bibitem{LSND} Aguilar A et al. (LSND Collaboration) 2001 {\it  Phys. Rev. D} {\bf 64} 112007
\bibitem{MiniBooNE} Aguilar A et al. (MiniBooNE Collaboration) 2013 {\it Phys. Rev. Lett.} {\bf 110} 161801
\bibitem{Giunti} Giunti C et al. 2013 {\it Phys. Rev. D} {\bf 88} 073008
\bibitem{SOX} Bellini G et al. (Borexino/SOX Collaboration) 2013 {\it JHEP} {\bf 08}  038
\bibitem{SOX_paralel} Vivier M for Borexino/SOX Collaboration: SOX: short baseline neutrino oscillations with Borexino. Parallel talk at TAUP 2015, Torino, Italy.


\end{thebibliography}
\end{document}